# A neglected topic in relativistic electrodynamics: transformation of electromagnetic integrals

Oleg D. Jefimenko
Physics Department, West Virginia University
P.O. Box 6315
Morgantown, WV 26506-6315

**Abstract**

Although relativistic electrodynamics is more than 100 year old, there is one neglected topic in its presentation and application: relativistic transformations of electromagnetic integrals. Whereas in theoretical and applied electrodynamics electric and magnetic fields are mainly expressed in terms of integrals over charge and current distributions, relativistic transformations are traditionally applied to point charges and elementary currents. The purpose of this paper is to show that relativistic transformations can be easily applied to electromagnetic integrals and that relativistic transformation of such integrals constitutes a valuable method for solving a variety of electromagnetic problems involving moving electromagnetic system. Six illustrative examples on relativistic transformation of electromagnetic integrals, including transformation of Maxwell's equations in their integral form, are presented in this paper.

**I. Introduction**

Relativistic transformations of electromagnetic quantities is an important element of relativistic electrodynamics. However, in standard presentations of relativistic electrodynamics such transformations are limited to electromagnetic quantities associated with "relativistic particles," which are usually rapidly moving point charges. On the other hand, electromagnetic theory deals mostly with electric and magnetic fields associated with charge and current distributions. These fields are usually expressed in terms of electromagnetic integrals. Quite clearly, manipulating electromagnetic integrals by means of relativistic transformations should be a useful extension of relativistic electrodynamics and should be preferably included in standard courses on relativistic electromagnetism.

This paper introduces six illustrative examples on relativistic transformations of electromagnetic integrals. In all six examples two inertial reference frames are used: a stationary reference frame $\Sigma$ ("laboratory"), represented by the Cartesian axes $x$, $y$, $z$, and an inertial reference frame $\Sigma'$, represented by the Cartesian axes $x'$, $y'$, $z'$, moving with velocity $\mathbf{v} = v\mathbf{i}$ relative to $\Sigma$ in the positive direction of their common $x$, $x'$ axis.

## II. Illustrative examples on relativistic transformation of electromagnetic integrals

***Example 1.*** The electric scalar potential of a stationary charge distribution $\rho$ is represented by the well-known equation

$$\varphi = \frac{1}{4\pi\varepsilon_0} \int \frac{\rho}{r} dV \ . \tag{1}$$

Using relativistic transformation equations, convert Eq. (1) into the integral for the magnetic vector potential produced by a charge distribution moving with constant velocity $\mathbf{v} = v\,\mathbf{i}$.

Let a charge distribution $\rho'$ be at rest in the moving reference frame $\Sigma'$. The electric potential $\varphi'$ produced by $\rho'$ in this reference frame is given by Eq. (1) with $\varphi$, $\rho$, $r$, and $dV$ replaced by the corresponding primed quantities:

$$\varphi' = \frac{1}{4\pi\varepsilon_0} \int \frac{\rho'}{r'} dV' \ . \tag{2}$$

Observed from the laboratory (reference frame $\Sigma$), the charge distribution $\rho'$ moves with velocity $v$ along a line parallel to the $x$ axis. Like all moving charges, it creates a magnetic field. To find the associated magnetic vector potential, we transform Eq. (2) by using appropriate transformation equations listed in the Appendix (these equations are identified by the prefix "A"). However, first we express $r'$ and $dV'$ appearing in Eq. (2) in terms of $x'$, $y'$, and $z'$:

$$\varphi' = \frac{1}{4\pi\varepsilon_0} \iiint \frac{\rho'}{(x'^2 + y'^2 + z'^2)^{1/2}} dx'dy'dz' \ . \tag{3}$$

Since we are free to chose the time of observation in the laboratory, we chose $t = 0$ for simplicity.[1] By Eqs. (A4), (A2) and (A3) we then have

$$dx' = d(\gamma dx),\ dy' = dy,\ dz' = dz \ . \tag{4}$$

By Eq. (A17) (noting that there is no current in $\Sigma'$ where the charge is stationary) we have

$$\rho' = \rho/\gamma \ . \tag{5}$$

By Eq. (A18) (noting that there is no magnetic vector potential in $\Sigma'$ where the charge is stationary) we have

$$A_x = \gamma v \varphi'/c^2 \ . \tag{6}$$

Substituting Eqs. (4)-(6) into Eq. (3), we obtain

$$A_x = \frac{\gamma v}{4\pi\varepsilon_0 c^2} \iiint \frac{\rho/\gamma}{[(\gamma x)^2 + y^2 + z^2]^{1/2}} d(\gamma x) dy dz = \frac{v}{4\pi\varepsilon_0 c^2} \iiint \frac{\rho}{[x^2 + (y^2 + z^2)/\gamma^2]^{1/2}} dx dy dz \ , \quad (7)$$

and since

$$1/\gamma^2 = 1 - v^2/c^2 \ , \quad (8)$$

we obtain upon simplifying the denominator of Eq. (7), replacing $dxdydz$ by $dV$, and replacing $1/\varepsilon_0 c^2$ by $\mu_0$,

$$A_x = \frac{\mu_0 v}{4\pi} \int \frac{\rho}{r[1 - (y^2 + z^2)v^2/r^2 c^2]^{1/2}} dV \ , \quad (9)$$

or

$$A_x = \frac{\mu_0 v}{4\pi} \int \frac{\rho}{r[1 - (v^2/c^2)\sin^2\theta]^{1/2}} dV \ , \quad (10)$$

where $\theta$ is the angle between the velocity vector **v** of the moving charge distribution and the radius vector **r** connecting $dV$ with the point of observation. For the $y$ and $z$ components of the vector potential we obtain from Eqs. (A19) and (A20)

$$A_y = A_z = 0 \ . \quad (11)$$

**Example 2.** The electric field of a stationary charge distribution $\rho$ is represented by the well-known equation

$$\mathbf{E} = \frac{1}{4\pi\varepsilon_0} \int \frac{\rho \mathbf{r}}{r^3} dV \ . \quad (12)$$

Applying relativistic transformation equations to Eq. (12), find the magnetic field produced by a charge distribution moving with constant velocity $\mathbf{v} = v\mathbf{i}$.

As in the preceding example, let a charge distribution $\rho'$ be at rest in the moving reference frame $\Sigma'$. Rewriting Eq. (12) in terms of its Cartesian components and prime coordinates, we have for the electric field produced by $\rho'$ in $\Sigma'$

$$E'_x = \frac{1}{4\pi\varepsilon_0} \iiint \frac{\rho' x'}{(x'^2 + y'^2 + z'^2)^{3/2}} dx' dy' dz' \ , \quad (13)$$

$$E'_y = \frac{1}{4\pi\varepsilon_0} \iiint \frac{\rho' y'}{(x'^2 + y'^2 + z'^2)^{3/2}} dx' dy' dz' \ , \quad (14)$$

$$E'_z = \frac{1}{4\pi\varepsilon_0} \iiint \frac{\rho' z'}{(x'^2 + y'^2 + z'^2)^{3/2}} dx' dy' dz' \ . \quad (15)$$

For the time of observation in the laboratory we chose as before $t = 0$, so that Eq. (4) applies again. Also, since there is no current in $\Sigma'$, Eq. (5) applies. By Eqs. (A8)-(A10) (noting that there is no magnetic field in $\Sigma'$ where the charge is stationary) we have

$$B_x = 0 \;, \tag{16}$$

$$B_y = -\gamma v E'_z / c^2 \;, \tag{17}$$

$$B_z = \gamma v E'_y / c^2 \;. \tag{18}$$

Substituting Eqs. (4), (5), (17), and (18) into Eqs. (15) and (14), we obtain

$$B_y = -\frac{\gamma v}{4\pi\varepsilon_0 c^2} \iiint \frac{(\rho/\gamma) z}{(\gamma^2 x^2 + y^2 + z^2)^{3/2}} d(\gamma x) dy dz \;, \tag{19}$$

$$B_z = \frac{\gamma v}{4\pi\varepsilon_0 c^2} \iiint \frac{(\rho/\gamma) y}{(\gamma^2 x^2 + y^2 + z^2)^{3/2}} d(\gamma x) dy dz \;. \tag{20}$$

Rewriting Eq. (16) and simplifying Eqs. (19) and (20) just as we simplified Eq. (7) in Example 1, we obtain for the magnetic field produced by a moving charge distribution

$$B_x = 0 \;, \tag{21}$$

$$B_y = -\frac{\mu_0 v}{4\pi} \int \frac{\rho(1 - v^2/c^2) z}{r^3 [1 - (v^2/c^2)\sin^2\theta]^{3/2}} dV \;, \tag{22}$$

$$B_z = \frac{\mu_0 v}{4\pi} \int \frac{\rho(1 - v^2/c^2) y}{r^3 [1 - (v^2/c^2)\sin^2\theta]^{3/2}} dV \;. \tag{23}$$

***Example 3.*** Show that Gauss's law for electric fields

$$\oint \mathbf{D} \cdot d\mathbf{S} = \int \rho dV \tag{24}$$

is invariant under relativistic transformations.

First we apply Gauss's theorem of vector analysis to Eq. (24), transforming it into

$$\int \nabla \cdot \mathbf{D} dV = \int \rho dV \;. \tag{25}$$

Expressing $\nabla \cdot \mathbf{D}$ in terms of its Cartesian components and making use of the relation $\mathbf{D} = \varepsilon_0 \mathbf{E}$, we then write Eq. (25) as

$$\varepsilon_0 \iiint \left( \frac{\partial E_x}{\partial x} + \frac{\partial E_y}{\partial y} + \frac{\partial E_z}{\partial z} \right) dx\,dy\,dz = \iiint \rho\,dx\,dy\,dz \ . \tag{26}$$

Let us now transform Eq. (26) to the moving reference frame $\Sigma'$. Let the time of observation in $\Sigma'$ be $t' = 0$. By Eqs. (A1)-(A3) we then have

$$dx = d(\gamma x'), \ dy = dy', \ dz = dz' \ . \tag{27}$$

Using Eq. (27) to replace $dx$, $dy$ and $dz$ in Eq. (26) by $dx'$, $dy'$ and $dz'$, and using Eqs. (A21), (A24) and (A25) to replace the derivatives with respect to $x$, $y$, $z$ by those with respect to $x'$, $y'$, $z'$, we have

$$\varepsilon_0 \iiint \left( \frac{1}{\gamma}\frac{\partial E_x}{\partial x'} - \frac{v}{c^2}\frac{\partial E_x}{\partial t} + \frac{\partial E_y}{\partial y'} + \frac{\partial E_z}{\partial z'} \right) d(\gamma x')\,dy'\,dz' = \iiint \rho\,d(\gamma x')\,dy'\,dz' \ . \tag{28}$$

Replacing now $E_x$, $E_y$ and $E_z$ in Eq. (28) (except in the derivative with respect to time) by $E'_x$, $E'_y$, $B_z$, $E'_z$, and $B_y$ by using Eqs. (A5), (A12) and (A13), and replacing $\rho$ by $\rho'$ and $J_x$ by using Eq. (A16), we obtain

$$\varepsilon_0 \iiint \left( \frac{1}{\gamma}\frac{\partial E'_x}{\partial x'} - \frac{v}{c^2}\frac{\partial E_x}{\partial t} + \frac{1}{\gamma}\frac{\partial E'_y}{\partial y'} + v\frac{\partial B_z}{\partial y'} + \frac{1}{\gamma}\frac{\partial E'_z}{\partial z'} - v\frac{\partial B_y}{\partial z'} \right) d(\gamma x')\,dy'\,dz' \tag{29}$$

$$= \iiint \left( \frac{1}{\gamma}\rho' + \frac{v}{c^2}J_x \right) d(\gamma x')\,dy'\,dz' \ ,$$

or

$$\varepsilon_0 \iiint \left( \frac{\partial E'_x}{\partial x'} + \frac{\partial E'_y}{\partial y'} + \frac{\partial E'_z}{\partial z'} \right) dx'\,dy'\,dz' + \varepsilon_0 v \iiint \left( \frac{\partial B_z}{\partial y'} - \frac{\partial B_y}{\partial z'} - \frac{1}{\varepsilon_0 c^2}J_x - \frac{1}{c^2}\frac{\partial E_x}{\partial t} \right) d(\gamma x')\,dy'\,dz' \tag{30}$$

$$= \iiint \rho'\,dx'\,dy'\,dz' \ .$$

Placing $\varepsilon_0$ under the integral signs and using $\varepsilon_0 \mathbf{E} = \mathbf{D}$ and $\varepsilon_0 \mathbf{B} = \varepsilon_0 \mu_0 \mathbf{H} = (1/c^2)\mathbf{H}$, we have

$$\iiint\left(\frac{\partial D'_x}{\partial x'}+\frac{\partial D'_y}{\partial y'}+\frac{\partial D'_z}{\partial z'}\right)dx'dy'dz' + v\iiint\left(\frac{\partial H_z}{c^2\partial y'}-\frac{\partial H_y}{c^2\partial z'}-\frac{1}{c^2}J_x-\frac{1}{c^2}\frac{\partial D_x}{\partial t}\right)d(\gamma x')dy'dz'$$

$$=\iiint\rho'dx'dy'dz' ,\tag{31}$$

or

$$\int\nabla'\cdot\mathbf{D}'dV' + \frac{v}{c^2}\iiint\left(\frac{\partial H_z}{\partial y'}-\frac{\partial H_y}{\partial z'}-J_x-\frac{\partial D_x}{\partial t}\right)d(\gamma x')dy'dz' = \int\rho'dV' .\tag{32}$$

However, by Eq. (27), $d(\gamma x')dy'dz' = dxdydz$, by Eqs. (A24) and (A25), $\partial/\partial y' = \partial/\partial y$, $\partial/\partial z' = \partial/\partial z$, and therefore, by Eq. (A31) the triple integral in Eq. (32) is zero. We thus obtain

$$\int\nabla'\cdot\mathbf{D}'dV' = \int\rho'dV'\tag{33}$$

and hence, by Gauss's theorem of vector analysis,

$$\oint\mathbf{D}'\cdot d\mathbf{S}' = \int\rho'dV' .\tag{34}$$

Thus our relativistic transformations from $\Sigma$ to $\Sigma'$ have not changed the form of Eq. (24). Hence Gauss's law for electric fields is invariant under relativistic transformations.

*Example 4.* Show that Gauss's law for magnetic fields

$$\oint\mathbf{B}\cdot d\mathbf{S} = 0\tag{35}$$

is invariant under relativistic transformations.

First we apply Gauss's theorem of vector analysis to Eq. (35), transforming it into

$$\int\nabla\cdot\mathbf{B}dV = 0 .\tag{36}$$

Expressing $\nabla\cdot\mathbf{B}$ in terms of its Cartesian components, we then write Eq. (36) as

$$\iiint\left(\frac{\partial B_x}{\partial x}+\frac{\partial B_y}{\partial y}+\frac{\partial B_z}{\partial z}\right)dxdydz = 0 .\tag{37}$$

Let us now transform Eq. (37) to the moving reference frame $\Sigma'$. As before, we choose $t' = 0$ for the time of observation in $\Sigma'$. Using Eq. (27) to replace $dx$, $dy$ and $dz$ in Eq. (37) by $dx'$, $dy'$ and $dz'$, and using Eqs. (A21), (A24) and (A25) to replace the derivatives with respect to $x$, $y$, $z$ by those with respect to $x'$, $y'$, $z'$, we have

$$\iiint \left( \frac{1}{\gamma} \frac{\partial B_x}{\partial x'} - \frac{v}{c^2} \frac{\partial B_x}{\partial t} + \frac{\partial B_y}{\partial y'} + \frac{\partial B_z}{\partial z'} \right) d(\gamma x') dy' dz' = 0 . \tag{38}$$

Replacing now $B_x$, $B_y$ and $B_z$ in Eq. (38) (except in the derivative with respect to time) by $B'_x$, $B'_y$, $E_z$, $B'_z$, and $E_y$ by means of Eqs. (A8), (A14) and (A15), we obtain

$$\iiint \left( \frac{1}{\gamma} \frac{\partial B'_x}{\partial x'} - \frac{v}{c^2} \frac{\partial B_x}{\partial t} + \frac{1}{\gamma} \frac{\partial B'_y}{\partial y'} - \frac{v}{c^2} \frac{\partial E_z}{\partial y'} + \frac{1}{\gamma} \frac{\partial B'_z}{\partial z'} + \frac{v}{c^2} \frac{\partial E_y}{\partial z'} \right) d(\gamma x') dy' dz' = 0 , \tag{39}$$

or

$$\iiint \left( \frac{\partial B'_x}{\partial x'} + \frac{\partial B'_y}{\partial y'} + \frac{\partial B'_z}{\partial z'} \right) dx' dy' dz' - \frac{v}{c^2} \iiint \left( \frac{\partial E_z}{\partial y'} - \frac{\partial E_y}{\partial z'} + \frac{\partial B_x}{\partial t} \right) d(\gamma x') dy' dz' = 0 , \tag{40}$$

and therefore

$$\int \nabla' \cdot \mathbf{B}' dV' - \frac{v}{c^2} \iiint \left( \frac{\partial E_z}{\partial y'} - \frac{\partial E_y}{\partial z'} + \frac{\partial B_x}{\partial t} \right) d(\gamma x') dy' dz' = 0 . \tag{41}$$

However, by Eq. (27), $d(\gamma x') dy' dz' = dx dy dz$, by Eqs. (A24) and (A25), $\partial/y' = \partial/y$, $\partial/z' = \partial/z$, and therefore, by Eq. (A36), the triple integral in Eq. (41) is zero. We thus obtain

$$\int \nabla' \cdot \mathbf{B}' dV' = 0 \tag{42}$$

and hence, by Gauss's theorem of vector analysis,

$$\oint \mathbf{B}' \cdot d\mathbf{S}' = 0 \tag{43}$$

Except for the primes, Eq. (43) is the same as Eq. (35). Hence Gauss's law for magnetic fields is invariant under relativistic transformations

*Example 5.* Show that Faraday's law

$$\oint \mathbf{E} \cdot d\mathbf{l} = -\frac{\partial}{\partial t} \int \mathbf{B} \cdot d\mathbf{S} \tag{44}$$

is invariant under relativistic transformations.

First we apply Stokes's theorem of vector analysis to Eq. (44), transforming it into

$$\int (\nabla \times \mathbf{E}) \cdot d\mathbf{S} = -\int \frac{\partial}{\partial t} \mathbf{B} \cdot d\mathbf{S} \ . \tag{45}$$

In terms of its Cartesian components Eq. (45) is

$$\iint \left( \frac{\partial E_z}{\partial y} - \frac{\partial E_y}{\partial z} \right) dydz + \iint \left( \frac{\partial E_x}{\partial z} - \frac{\partial E_z}{\partial x} \right) dzdx + \iint \left( \frac{\partial E_y}{\partial x} - \frac{\partial E_x}{\partial y} \right) dxdy$$

$$= -\iint \frac{\partial B_x}{\partial t} dydz - \iint \frac{\partial B_y}{\partial t} dzdx - \iint \frac{\partial B_z}{\partial t} dxdy. \tag{46}$$

Let us now transform Eq. (46) to the moving reference frame Σ′. As before, we choose $t'$ = 0 for the time of observation in Σ′. Using Eq. (27) to replace $dx$, $dy$ and $dz$ in Eq. (46) by $dx'$, $dy'$ and $dz'$, and using Eqs. (A24) and (A25) to replace the derivatives with respect to $y$ and $z$ by those with respect to $y'$ and $z'$, we have

$$\iint \left( \frac{\partial E_z}{\partial y'} - \frac{\partial E_y}{\partial z'} \right) dy'dz' + \iint \left( \frac{\partial E_x}{\partial z'} - \frac{\partial E_z}{\partial x} \right) dz'd(\gamma x') + \iint \left( \frac{\partial E_y}{\partial x} - \frac{\partial E_x}{\partial y'} \right) d(\gamma x')dy'$$

$$= -\iint \frac{\partial B_x}{\partial t} dy'dz' - \iint \frac{\partial B_y}{\partial t} dz'd(\gamma x') - \iint \frac{\partial B_z}{\partial t} d(\gamma x')dy' \ . \tag{47}$$

Using now Eqs. (A5)-(A7) to replace $E_x$, $E_y$, $E_z$ in Eq. (47) by the corresponding primed quantities (except in the derivatives with respect to $x$), using Eq. (A8) to replace $B_x$ by $B'_x$, and using Eq. (A22) to replace $\partial/\partial t$ in the first term on the right of Eq. (47) by the corresponding primed quantities, we can write Eq. (47) as

$$\iint \left( \gamma \frac{\partial E'_z}{\partial y'} - \gamma w \frac{\partial B'_y}{\partial y'} - \gamma \frac{\partial E'_y}{\partial z'} - \gamma w \frac{\partial B'_z}{\partial z'} \right) dy'dz' + \iint \left( \frac{\partial E'_x}{\partial z'} - \frac{\partial E_z}{\partial x} \right) dz'd(\gamma x') + \iint \left( \frac{\partial E_y}{\partial x} - \frac{\partial E'_x}{\partial y'} \right) d(\gamma x')dy'$$

$$= -\iint \left( \gamma \frac{\partial B'_x}{\partial t'} - \gamma w \frac{\partial B'_x}{\partial x'} \right) dy'dz' - \iint \frac{\partial B_y}{\partial t} dz'd(\gamma x') - \iint \frac{\partial B_z}{\partial t} d(\gamma x')dy' \ . \tag{48}$$

According to Eq. (42), taking into account that Eq. (42) holds for any volume of integration and that therefore $\nabla' \cdot \mathbf{B}' = 0$, the terms with $\partial B'_x/\partial x'$, $\partial B'_y/\partial y'$ and $\partial B'_z/\partial z'$ in Eq. (48) vanish, so that Eq. (48) simplifies to

$$\iint\left(\gamma\frac{\partial E'_z}{\partial y'}-\gamma\frac{\partial E'_y}{\partial z'}\right)dy'dz' + \iint\left(\frac{\partial E'_x}{\partial z'}-\frac{\partial E_z}{\partial x}\right)dz'd(\gamma x') + \iint\left(\frac{\partial E_y}{\partial x}-\frac{\partial E'_x}{\partial y'}\right)d(\gamma x')dy'$$

$$=-\iint\gamma\frac{\partial B'_x}{\partial t'}dy'dz' - \iint\frac{\partial B_y}{\partial t}dz'd(\gamma x') - \iint\frac{\partial B_z}{\partial t}d(\gamma x')dy' \ . \tag{49}$$

Using Eqs. (A7), (A6), (A9), and (A10), we can write Eq. (49) as

$$\iint\left(\gamma\frac{\partial E'_z}{\partial y'}-\gamma\frac{\partial E'_y}{\partial z'}\right)dy'dz' + \iint\left[\frac{\partial E'_x}{\partial z'}-\gamma\frac{\partial(E'_z-vB'_y)}{\partial x}\right]dz'd(\gamma x') + \iint\left[\gamma\frac{\partial(E'_y+vB'_z)}{\partial x}-\frac{\partial E'_x}{\partial y'}\right]d(\gamma x')dy'$$

$$=-\iint\gamma\frac{\partial B'_x}{\partial t'}dy'dz' - \iint\gamma\frac{\partial(B'_y-vE'_z/c^2)}{\partial t}dz'd(\gamma x') - \iint\gamma\frac{\partial(B'_z+vE'_y/c^2)}{\partial t}d(\gamma x')dy' \ . \tag{50}$$

Dividing Eq. (50) by $\gamma$ and rearranging, we obtain

$$\iint\left(\frac{\partial E'_z}{\partial y'}-\frac{\partial E'_y}{\partial z'}\right)dy'dz'$$

$$+\iint\left[\frac{\partial E'_x}{\partial z'}-\gamma\left(\frac{\partial E'_z}{\partial x}+\frac{v}{c^2}\frac{\partial E'_z}{\partial t}\right)\right]dz'dx' + \iint\left[\gamma\left(\frac{\partial E'_y}{\partial x}+\frac{v}{c^2}\frac{\partial E'_y}{\partial t}\right)-\frac{\partial E'_x}{\partial y'}\right]dx'dy' \tag{51}$$

$$=-\iint\frac{\partial B'_x}{\partial t'}dy'dz' - \iint\gamma\left(\frac{\partial B'_y}{\partial t}+v\frac{\partial B'_y}{\partial x}\right)dz'dx' - \iint\gamma\left(\frac{\partial B'_z}{\partial t}+v\frac{\partial B'_z}{\partial x}\right)dx'dy' \ ,$$

which, by Eqs. (A21) and (A23), is

$$\iint\left(\frac{\partial E'_z}{\partial y'}-\frac{\partial E'_y}{\partial z'}\right)dy'dz' + \iint\left(\frac{\partial E'_x}{\partial z'}-\frac{\partial E'_z}{\partial x'}\right)dz'dx' + \iint\left(\frac{\partial E'_y}{\partial x'}-\frac{\partial E'_x}{\partial y'}\right)dx'dy'$$

$$=-\iint\frac{\partial B'_x}{\partial t'}dy'dz' - \iint\frac{\partial B'_y}{\partial t'}dz'dx' - \iint\frac{\partial B'_z}{\partial t'}dx'dy' \ , \tag{52}$$

or, in vector notation,

$$\int(\nabla'\times\mathbf{E}')\cdot d\mathbf{S}' = -\int\frac{\partial}{\partial t'}\mathbf{B}'\cdot d\mathbf{S}' , \tag{53}$$

and, by Stokes's theorem of vector analysis,

$$\oint \mathbf{E}' \cdot d\mathbf{l}' = -\frac{\partial}{\partial t'} \int \mathbf{B}' \cdot d\mathbf{S}'. \tag{54}$$

Except for the primes, Eq. (54) is the same as Eq. (44). Hence Faraday's law Eq. (44) is invariant under relativistic transformations.

***Example 6.*** Show that the equation

$$\oint \mathbf{H} \cdot d\mathbf{l} = \int \left( \mathbf{J} + \frac{\partial}{\partial t} \mathbf{D} \right) \cdot d\mathbf{S} \tag{55}$$

is invariant under relativistic transformations.

First we apply Stokes's theorem of vector analysis to Eq. (55), transforming it into

$$\int (\nabla \times \mathbf{H}) \cdot d\mathbf{S} = \int \left( \mathbf{J} + \frac{\partial}{\partial t} \mathbf{D} \right) \cdot d\mathbf{S} . \tag{56}$$

In terms of its Cartesian components Eq. (56) is

$$\iint \left( \frac{\partial H_z}{\partial y} - \frac{\partial H_y}{\partial z} \right) dy dz + \iint \left( \frac{\partial H_x}{\partial z} - \frac{\partial H_z}{\partial x} \right) dz dx + \iint \left( \frac{\partial H_y}{\partial x} - \frac{\partial H_x}{\partial y} \right) dx dy$$

$$= \iint \left( J_x + \frac{\partial D_x}{\partial t} \right) dy dz + \iint \left( J_y + \frac{\partial D_y}{\partial t} \right) dz dx + \iint \left( J_z + \frac{\partial D_z}{\partial t} \right) dx dy. \tag{57}$$

Let us now transform Eq. (57) to the moving reference frame $\Sigma'$. As before, we choose $t' = 0$ for the time of observation in $\Sigma'$. Using Eq. (27) to replace $dx$, $dy$ and $dz$ in Eq. (57) by $dx'$, $dy'$ and $dz'$, using Eqs. (A24) and (A25) to replace the derivatives with respect to $y$ and $z$ by those with respect to $y'$ and $z'$, and using the relation $\mathbf{D} = \varepsilon_0 \mathbf{E}$, we have

$$\iint \left( \frac{\partial H_z}{\partial y'} - \frac{\partial H_y}{\partial z'} \right) dy' dz' + \iint \left( \frac{\partial H_x}{\partial z'} - \frac{\partial H_z}{\partial x} \right) dz' d(\gamma x') + \iint \left( \frac{\partial H_y}{\partial x} - \frac{\partial H_x}{\partial y'} \right) d(\gamma x') dy'$$

$$= \iint \left( J_x + \varepsilon_0 \frac{\partial E_x}{\partial t} \right) dy' dz' + \iint \left( J_y + \varepsilon_0 \frac{\partial E_y}{\partial t} \right) dz' d(\gamma x') + \iint \left( J_z + \varepsilon_0 \frac{\partial E_z}{\partial t} \right) d(\gamma x') dy' . \tag{58}$$

Remembering that $\mathbf{B} = \mu_0 \mathbf{H}$, using Eqs. (A8)-(A10) to replace $H_x$, $H_y$, $H_z$ in Eq. (58) (except in the derivatives with respect to $x$) by the corresponding primed quantities, using Eq. (A26) to replace $J_x$ by $J'_x$ and $\rho'$, using Eq. (A5) to replace $E_x$ by $E'_x$, and using Eq. (A22) to replace $\partial/\partial t$ in the first term on the right of Eq. (58) by the corresponding primed quantities, we can write Eq. (58) as

$$\iint \left( \gamma \frac{\partial H'_z}{\partial y'} + \gamma v \frac{\partial E'_y}{\mu_0 c^2 \partial y'} - \gamma \frac{\partial H'_y}{\partial z'} + \gamma v \frac{\partial E'_z}{\mu_0 c^2 \partial z'} \right) dy' dz'$$

$$+ \iint \left( \frac{\partial H'_x}{\partial z'} - \frac{\partial H_z}{\partial x} \right) dz' d(\gamma x') + \iint \left( \frac{\partial H_y}{\partial x} - \frac{\partial H'_x}{\partial y'} \right) d(\gamma x') dy' \quad (59)$$

$$= \iint \left[ \gamma (J'_x + v\rho') + \varepsilon_0 \gamma \left( \frac{\partial E'_x}{\partial t'} - v \frac{\partial E'_x}{\partial x'} \right) \right] dy' dz' + \iint \left( J'_y + \varepsilon_0 \frac{\partial E_y}{\partial t} \right) dz' d(\gamma x') + \iint \left( J'_z + \varepsilon_0 \frac{\partial E_z}{\partial t} \right) d(\gamma x') dy' \,.$$

According to Eq. (33), taking into account that $1/\mu_0 c^2 = \varepsilon_0$, taking into account that Eq. (33) holds for any volume of integration and that therefore $\nabla' \cdot \mathbf{D}' = \rho'$, and also taking into account that $\mathbf{D}' = \varepsilon_0 \mathbf{E}'$, the terms with $\partial E'_x / \partial x'$, $\partial E'_y / \partial y'$, $\partial E'_z / \partial z'$ and $\rho$ in Eq. (59) vanish, so that Eq. (59) simplifies to

$$\iint \left( \gamma \frac{\partial H'_z}{\partial y'} - \gamma \frac{\partial H'_y}{\partial z'} \right) dy' dz' + \iint \left( \frac{\partial H'_x}{\partial z'} - \frac{\partial H_z}{\partial x} \right) dz' d(\gamma x') + \iint \left( \frac{\partial H_y}{\partial x} - \frac{\partial H'_x}{\partial y'} \right) d(\gamma x') dy'$$

$$= \iint \left[ \gamma J'_x + \varepsilon_0 \gamma \frac{\partial E'_x}{\partial t'} \right] dy' dz' + \iint \left( J'_y + \varepsilon_0 \frac{\partial E_y}{\partial t} \right) dz' d(\gamma x') + \iint \left( J'_z + \varepsilon_0 \frac{\partial E_z}{\partial t} \right) d(\gamma x') dy' \,. \quad (60)$$

Dividing Eq. (60) by $\gamma$, taking into account that $\mathbf{B}' = \mu_0 \mathbf{H}'$, and using Eqs. (A10), (A9), (A6), and (A7), we can write Eq. (60) as

$$\iint \left( \frac{\partial H'_z}{\partial y'} - \frac{\partial H'_y}{\partial z'} \right) dy' dz'$$

$$+ \iint \left[ \frac{\partial H'_x}{\partial z'} - \gamma \frac{\partial (H'_z + v E'_y / \mu_0 c^2)}{\partial x} \right] dz' dx' + \iint \left[ \gamma \frac{\partial (H'_y - v E'_z / \mu_0 c^2)_y}{\partial x} - \frac{\partial H'_x}{\partial y'} \right] dx' dy' \quad (61)$$

$$= \iint \left( J'_x + \varepsilon_0 \frac{\partial E'_x}{\partial t'} \right) dy' dz' + \iint \left[ J'_y + \varepsilon_0 \gamma \frac{\partial (E'_y + v B'_z)}{\partial t} \right] dz' dx' + \iint \left[ J'_z + \varepsilon_0 \gamma \frac{\partial (E'_z - v B'_y)}{\partial t} \right] dx' dy' \,.$$

Using the relations $\varepsilon_0 \mathbf{B} = \mathbf{H}/c^2$ and $1/\mu_0 c^2 = \varepsilon_0$, we next rearrange Eq. (61) into

$$\iint\left(\frac{\partial H'_z}{\partial y'} - \frac{\partial H'_y}{\partial z'}\right)dy'dz'$$

$$+ \iint\left[\frac{\partial H'_x}{\partial z'} - \gamma\left(\frac{\partial H'_z}{\partial x} + \frac{v}{c^2}\frac{\partial H'_z}{\partial t}\right)\right]dz'dx' + \iint\left[\gamma\left(\frac{\partial H'_y}{\partial x} + \frac{v}{c^2}\frac{\partial H'_y}{\partial t}\right) - \frac{\partial H'_x}{\partial y'}\right]dx'dy' \quad (62)$$

$$= \iint\left(J'_x + \varepsilon_0\frac{\partial E'_x}{\partial t'}\right)dy'dz' + \iint\left[J'_y + \varepsilon_0\gamma\left(\frac{\partial E'_y}{\partial t} + v\frac{\partial E'_y}{\partial x}\right)\right]dz'dx' + \iint\left[J'_z + \varepsilon_0\gamma\left(\frac{\partial E'_z}{\partial t} + v\frac{\partial E'_z}{\partial x}\right)\right]dx'dy' \quad.$$

By Eqs. (A21) and (A23), Eq. (62) can be written as

$$\iint\left(\frac{\partial H'_z}{\partial y'} - \frac{\partial H'_y}{\partial z'}\right)dy'dz' + \iint\left(\frac{\partial H'_x}{\partial z'} - \frac{\partial H'_z}{\partial x'}\right)dz'dx' + \iint\left(\frac{\partial H'_y}{\partial x'} - \frac{\partial H'_x}{\partial y'}\right)dx'dy'$$

$$= \iint\left(J'_x + \varepsilon_0\frac{\partial E'_x}{\partial t'}\right)dy'dz' + \iint\left(J'_y + \varepsilon_0\frac{\partial E'_y}{\partial t'}\right)dz'dx' + \iint\left(J'_z + \varepsilon_0\frac{\partial E'_z}{\partial t'}\right)dx'dy' \quad, \quad (63)$$

or, in vector notation, remembering that $\varepsilon_0 \mathbf{E} = \mathbf{D}$, as

$$\int(\nabla' \times \mathbf{H'}) \cdot d\mathbf{S'} = \int\left(\mathbf{J'} + \frac{\partial}{\partial t'}\mathbf{D'}\right) \cdot d\mathbf{S'}, \quad (64)$$

and, by Stokes's theorem of vector analysis, as

$$\oint \mathbf{H'} \cdot d\mathbf{l'} = \int\left(\mathbf{J'} + \frac{\partial \mathbf{D'}}{\partial t'}\right) \cdot d\mathbf{S'}. \quad (65)$$

Except for the primes, Eq. (65) is the same as Eq. (55). Hence Eq. (55) is invariant under relativistic transformations.

**III. Discussion**

Examples 1 and 2 show the remarkable power and utility of relativistic transformation equations. One cannot help but to be impressed by the fact that applying relativistic transformations to equations for the electrostatic potential and field we obtain equations for the magnetic vector potential and field,[2] although at first sight there appears to be no connection whatsoever between electrostatic and magnetic quantities. Of course, the primary purpose of these examples in this paper was to demonstrate relativistic operations with electromagnetic integrals, and the very impressive outcome of these examples should not obscure their primary purpose. Still, one cannot ignore the fact that obtaining the integrals for the magnetic potential and field without using relativistic transformations is a rather difficult process.[3,4]

Examples 3-4, although not quite so impressive as Examples 1 and 2, are nevertheless of considerable significance. The reader has undoubtedly noticed that Eqs. (24), (35), (44), and (55) are Maxwell's equations in their integral form. In Einstein's famous 1905 paper[5] and in all early publications on relativistic electrodynamics, only the invariance of the differential form of Maxwell's equations under relativistic transformations was discussed or demonstrated. As far as this author knows, all later and the most recent publications[6] on relativistic electromagnetism have adhered to this custom. As a result, the students of relativistic electromagnetism receive an unavoidable impression that relativistic transformations do not work with equations involving integrals, and, in particular, are not applicable to the integral form of Maxwell's equation. However, as it is shown in this paper, relativistic transformations work not only perfectly well with electromagnetic integrals, but yield very gratifying results specifically when applied to such integrals.

## APPENDIX

### (a) Relativistic transformation equations used in this paper[7]

$$x = \gamma(x' + vt') \tag{A1}$$
$$y = y' \tag{A2}$$
$$z = z' \tag{A3}$$
$$x' = \gamma(x - vt) \tag{A4}$$
$$E_x = E'_x \tag{A5}$$
$$E_y = \gamma(E'_y + vB'_z) \tag{A6}$$
$$E_z = \gamma(E'_z - vB'_y) \tag{A7}$$
$$B_x = B'_x \tag{A8}$$
$$B_y = \gamma(B'_y - vE'_z/c^2) \tag{A9}$$
$$B_z = \gamma(B'_z + vE'_y/c^2) \tag{A10}$$
$$t' = \gamma(t - vx/c^2) \tag{A11}$$
$$E_y = \frac{1}{\gamma}E'_y + vB_z \tag{A12}$$
$$E_z = \frac{1}{\gamma}E'_z - vB_y \tag{A13}$$
$$B_y = \frac{1}{\gamma}B'_y - vE_z/c^2 \tag{A14}$$
$$B_z = \frac{1}{\gamma}B'_z + vE_y/c^2 \tag{A15}$$
$$\rho = \frac{1}{\gamma}\rho' + (v/c^2)J_x \tag{A16}$$
$$\rho' = \frac{1}{\gamma}\rho - (v/c^2)J'_x \tag{A17}$$

$$A_x = \gamma[A'_x + (v/c^2)\varphi'] \tag{A18}$$

$$A_y = A'_y \tag{A19}$$

$$A_z = A'_z \tag{A20}$$

$$\frac{\partial}{\partial x} = \frac{1}{\gamma}\frac{\partial}{\partial x'} - \frac{v}{c^2}\frac{\partial}{\partial t} \tag{A21}$$

$$\frac{\partial}{\partial t} = \gamma\left(\frac{\partial}{\partial t'} - v\frac{\partial}{\partial x'}\right) \tag{A22}$$

$$\frac{\partial}{\partial t'} = \gamma\left(\frac{\partial}{\partial t} + v\frac{\partial}{\partial x}\right) \tag{A23}$$

$$\frac{\partial}{\partial y} = \frac{\partial}{\partial y'} \tag{A24}$$

$$\frac{\partial}{\partial z} = \frac{\partial}{\partial z'} \tag{A25}$$

$$J_x = \gamma(J'_x + v\rho') \tag{A26}$$

### (a) Supplementary calculations

In Example 3 we encounter the integral

$$\iint\left(\frac{\partial H_z}{\partial y} - \frac{\partial H_y}{\partial z} - J_x - \frac{\partial D_x}{\partial t}\right)dydz \ . \tag{A27}$$

Let us show that this integral is zero. Applying Stokes's theorem of vector analysis to Maxwell's equation

$$\oint \mathbf{H}\cdot d\mathbf{l} = \int\left(\mathbf{J} + \frac{\partial}{\partial t}\mathbf{D}\right)\cdot d\mathbf{S}, \tag{A28}$$

we have

$$\int(\nabla\times\mathbf{H})\cdot d\mathbf{S} = \int\left(\mathbf{J} + \frac{\partial}{\partial t}\mathbf{D}\right)\cdot d\mathbf{S} \ , \tag{A29}$$

or

$$\int\left(\nabla\times\mathbf{H} - \mathbf{J} - \frac{\partial}{\partial t}\mathbf{D}\right)\cdot d\mathbf{S} = 0 \ . \tag{A30}$$

Since Eq. (A30) holds for any surface of integration, the integrand in it is zero. And since the integrand is a vector, all its Cartesian components are zero. For the $x$ component of the integrand we then have

$$\iint \left( \frac{\partial H_z}{\partial y} - \frac{\partial H_y}{\partial z} - J_x - \frac{\partial D_x}{\partial t} \right) dydz = 0 \ . \tag{A31}$$

We shall also encounter the integral

$$\iint \left( \frac{\partial E_z}{\partial y} - \frac{\partial E_y}{\partial z} + \frac{\partial B_x}{\partial t} \right) dydz \ . \tag{A32}$$

Let us show that this integral is zero.

Applying Stokes's theorem of vector analysis to Faraday's law

$$\oint \mathbf{E} \cdot d\mathbf{l} = -\frac{\partial}{\partial t} \int \mathbf{B} \cdot d\mathbf{S}, \tag{A33}$$

we have

$$\int (\nabla \times \mathbf{E}) \cdot d\mathbf{S} = -\int \frac{\partial}{\partial t} \mathbf{B} \cdot d\mathbf{S} \ , \tag{A34}$$

or

$$\int \left( \nabla \times \mathbf{E} + \frac{\partial}{\partial t} \mathbf{B} \right) \cdot d\mathbf{S} = 0 \ . \tag{A35}$$

Since Eq. (A35) holds for any surface of integration, the integrand in it is zero. And since the integrand is a vector, all its Cartesian components are zero. For the *x* component of the integrand we then have

$$\iint \left( \frac{\partial E_z}{\partial y} - \frac{\partial E_y}{\partial z} + \frac{\partial B_x}{\partial t} \right) dydz = 0 \ . \tag{A36}$$

### References and Remarks

**1.** One may think that since the values of φ′, ρ′ and r′ in Σ′ correspond to some time of observation *t*′ = *constant,* they cannot be transformed to *t* = *constant* in Σ by Lorentz transformations, and therefore using *t* = 0 in Σ is in violation of the theory of relativity. However, by the principle of relativity, if a physical law does not depend on time in one inertial reference frame, then it (or its equivalent) cannot depend on time in any other inertial reference frame. Moreover, if a quantity does not explicitly appear in the formula under consideration, it is not subject to Lorentz transformation in the course of transforming this formula to a different reference frame. Therefore, since Eq. (2) in Σ′ is time-independent, and since *t* does not appear in Eq. (2) (the time of observation in Σ′ in unspecified), we are free to choose any convenient time of observation in Σ. Similarly, since the actual volume of integration in Eq. (2) in Σ′ is arbitrary (as long as it encloses the charge distribution under consideration), we need not transform it to some particular volume in $\Sigma$: More generally, if the domain of integration (volume, surface, contour) in one inertial reference frame is unspecified, it is not subject to Lorentz transformation in the course of transforming the corresponding integral to a different reference frame.

**2.** Observed that for low velocities, when $v^2/c^2$ can be neglected, Eqs. (10) and (11) reduce to the components of the better-known equation $\mathbf{A} = \dfrac{\mu_0}{4\pi} \int \dfrac{\mathbf{J}}{r} dV$, and Eqs. (21)-(23) similarly reduce to the components of $\mathbf{B} = \dfrac{\mu_0}{4\pi} \int \dfrac{\mathbf{J} \times \mathbf{r}}{r^3} dV$ with $\mathbf{J} = \rho\mathbf{v}$.

**3.** See, for example, M. Mason and W. Weaver, *The Electromagnetic Field*, (The University of Chicago Press, Chicago, Illinois, 1929) pp. 285-301.

**4.** For additional examples of transformation of electromagnetic integrals see O. D. Jefimenko, *Electromagnetic Retardation and Theory of Relativity*, 2nd ed.,(Electret Scientific, Star City, 2004) pp. 171-175 and 176-178.

**5.** A. Einstein, "Zur Elektrodynamik bewegter Körper," Ann. Phys. **17**, 891-921 (1905).

**6.** Typical examples are W. G. V. Rosser, *Interpretation of Classical Electromagnetism*, (Kluwer, Dordrecht, 1997) pp. 355-382 and J. D. Jackson, *Classical Electrodynamics*, 3rd ed., (Wiley, New York, 1999) pp. 514-578.

**7.** For the derivation of these equations see, for example, W. G. V. Rosser, *Classical Electromagnetism Via Relativity* (Plenum, New York, 1968) pp. 6, 151-158 and Ref. 4, pp. 150-158.